\documentstyle[iopconf1,epsf]{article}
\begin{document}
\title{Cosmic Strings and Structure Formation}
\author{Carsten van de Bruck\dag}
\affil{\dag\ Max--Planck Institut f\"ur Radioastronomie, Auf dem 
H\"ugel 69, D--53121 Bonn, Germany}
\beginabstract
We discuss structure formation in cosmic string theories. 
Emphasis is laid on the properties of the peculiar velocity field, 
non--gaussian features and clusters of galaxies. It is found that 
the predicted peculiar velocity field is quiet ($v<150$ km/s for 
$L>30 h^{-1}$ Mpc) for models 
with ($\Omega_0 h \leq 0.2$) which is in disagreement with 
the results from POTENT at large scales ($>30h^{-1}$ Mpc), consistent, 
however, with the analysis of the peculiar velocity field of spiral 
galaxies by Giovannelli et al. \cite{giov}. 
Using the Press--Schechter formalism we calculate 
the abundances of X--ray clusters. It is found that in CDM models a 
mass fluctuation $\sigma_8 = 0.6...0.7$ is needed with 
$\Omega_0 h =0.1...0.2$ in order to explain the current data. 
\endabstract

\section{Introduction}
An outstanding problem in cosmology is to explain the observed structures 
in the universe \cite{botun}. Two paradigms have been proposed: 
the inflationary cosmology and 
topological defects, which are well motivated by particle physics 
\cite{kolbturner,vilenkinshellard,brandenberger,hindmarsh}. 
Both could, at least in principle, be tested with observations of the 
microwave background anisotropies and with deep redshift surveys, because 
they make different predictions of the matter distribution and of the 
spectrum of the anisotropies of the cosmic microwave background radiation 
(CMBR). 

We discuss here the cosmic string theory, its prediction of the properties of 
the peculiar velocity field, the non-gaussian features 
in cosmic string models and the implications of non--gaussianity 
on the abundances of clusters of galaxies. We do not discuss power 
spectra and CMBR anisotropies and refer instead to the contribution 
by Richard Battye (these proceedings and references therein). 

The paper is organized as follows: in section 2 we describe the cosmic 
string network evolution on large scales, which is relevant for structure 
formation. In section 3 we discuss the properties of the 
peculiar velocity field and in section 4 the non--gaussian features 
in cosmic string theory. Our conclusions are given in section 5. 

We take the opportunity to include some points which could not be covered 
in the talk.

\begin{table}
\begin{center}
\begin{tabular}{rrrrrr}
\hline 
Model & $k$ & $\Omega_{0}$ & $\lambda_{0}$ & H$_{0}$ & log($R_{0}/R_{eq}$)\\
\hline
1     & +1& 0.014 & 1.08 & 90 & 2.66\\
\hline
2     & 0 & 1.0   & 0.0  & 60 & 4.16\\
\hline
3     & -1& 0.1   & 0.0  & 60 & 3.16\\
\hline
4     & 0 & 0.1   & 0.9  & 60 & 3.16\\
\hline
\end{tabular}
\caption{The four representative cosmological models. $k$ is the 
curvature parameter, $\Omega_0=8\pi G \rho/(3H_0^2)$ is the matter 
density parameter, $\lambda_{0}=\Lambda/(3H_0^2)$ is the cosmological term, 
$H_{0}$ is the Hubble parameter (in km s$^{-1}$Mpc$^{-1}$). $R_{0}$ and 
$R_{eq}$ are the scale factor at the present time and at matter--radiation 
equality, respectively.}
\end{center}
\end{table}

\section{Network evolution}
A network of cosmic strings, which are line--like objects, will 
originate in a phase transition if the vacuum manifold is 
non--simply connected. To know how cosmic strings could form structures, 
we need to understand the evolution of such a network. Therefore we 
discuss this in more detail in this section, where we focus on the 
case of local cosmic strings. 

Our approach is based on the modified ``one--scale'' model by 
Martins \& Shellard \cite{martinsshellard}. 
In this (phenomenological) model, the string network is characterized 
by a length scale $L$ and the RMS velocity $v_{\rm RMS}$ of the strings. 
The length scale $L$ is defined by
\begin{equation}
\rho_{\infty} = \frac{\mu}{L^2},
\end{equation}
where $\rho_{\infty}$ is the energy density contained in the long 
strings (i.e. the strings with curvature radius larger than the Hubble radius) 
and $\mu$ the energy per unit length on the string\footnote{$\mu$ is
related to the energy scale of the phase transition where the string network 
was created.}. The model can not only be 
applied to cosmic strings but also to line defects in the condensed matter 
context. The reader is referred to the paper by Martins \& Shellard, where 
further material and justifications of the equations are 
given \cite{martinsshellard}. We note that 
the equations are only exact in flat space--times. However, curvature 
effects can easily been incoporated. Our results presented below 
do not depend strongly on 
curvature effects. Curvature leads only to a faster release of energy of 
the network. Our results are in agreement with \cite{avelino1}.

\begin{figure}
\begin{center}
    \leavevmode
    \epsfxsize=8.9cm
   \epsffile{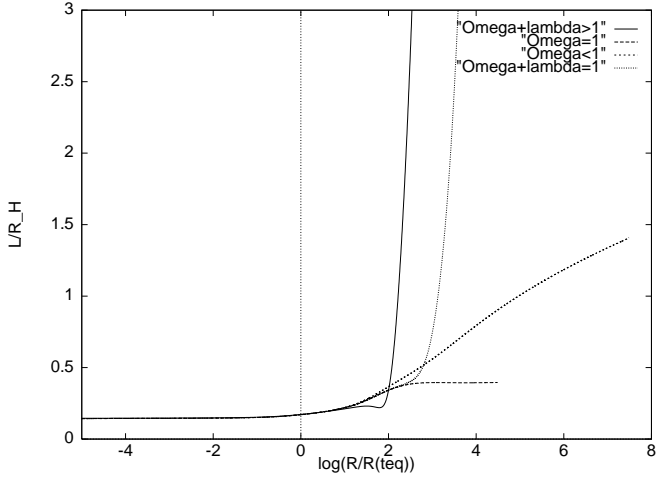}\vspace{0.5cm}
\caption{Behaviour of $L/R_H$ as a function of $log(R/R(t_{eq}))$, 
where $R(t_{eq})$ is the scale factor at the time of matter--radiation equality 
and $R_H$ is the (time--dependent) Hubble radius, adapted from 
\cite{vdbruck1}.}
\end{center}
\end{figure}

The equations, which determine the parameter $L$ and $v_{\rm RMS}$, are 
given by (we set the speed of light $c=1$):
\begin{equation}
\frac{dL}{dt} = H L(1 + v_{\rm RMS}^2) + \frac{\tilde{c} v_{\rm RMS}}{2}.
\end{equation}

 \begin{equation}
\frac{dv_{\rm RMS}}{dt} = (1 - v_{\rm RMS}^2)\left(\frac{\kappa}{r} 
- 2 H v_{\rm RMS} \right),
\end{equation}
where $\kappa$ is a parameter related to the small scale structure 
on the strings and $\tilde{c}$ is a parameter describing the energy loss 
due to loop formation. The expansion rate $H(t)$ of the universe 
obeys the Friedmann equation:
\begin{equation}
H^2(t) =\left(\frac{\dot R (t)}{R(t)}\right)^{2} = \frac{8 \pi G}{3}\rho(t) 
+ \frac{\Lambda}{3} - \frac{k}{R(t)^{2}},
\end{equation}
where $\rho=\rho_{\rm matter} + \rho_{\rm radiaton}$ is sum 
of the total matter density and radiation density, $\Lambda$ the cosmological 
constant, $k$ the curvature parameter and $R$ the scale factor. 

We solve the network equations for four cosmological models summarized
in the table 1 \cite{vdbruck1}. 
The models include an open model, a flat model with a 
cosmological constant, a model with a loitering phase (see the article 
by van de Bruck and Priester, these proceedings) and the 
Einstein--de Sitter model for comparison. The behaviour of the length 
scale $L$ is plotted in Fig.1 and Fig.2. 

\begin{figure}
\begin{center}
    \leavevmode
    \epsfxsize=8.9cm
    \epsffile{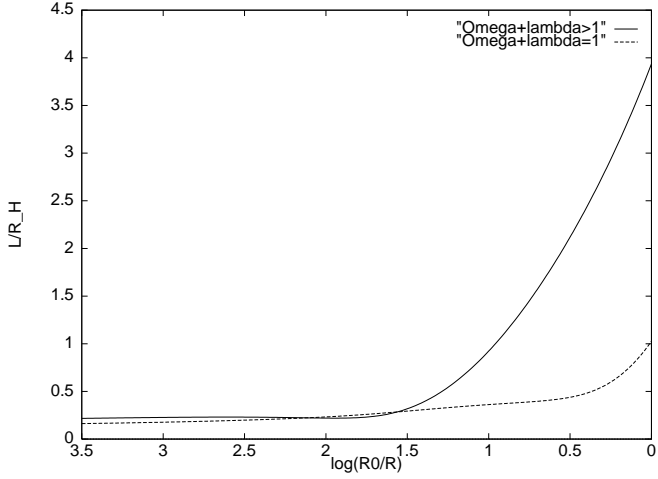}\vspace{0.5cm}
\caption{Behaviour of $L/R_H$ as a function of $log(R_0/R)$ for the 
models 1 and 4. $R_0$ is the scale factor at the present time 
and $R_H$ is the Hubble radius, see \cite{vdbruck1}.}
\end{center}
\end{figure}

It is usually argued that the cosmic string network evolves according 
to the scaling solution, which states that the ratio of the length $L$ 
to the Hubble radius $R_H (t) = 1/H(t)$ is 
constant: {\it scaling solution} $L/R_H = $ const.
We found that the scaling behaviour is always found in the radiation 
dominated epoch, but only reached in the matter dominated epoch 
in the Einstein--de Sitter model. In all other models, the scaling 
behaviour can not be found after the radiation dominated 
epoch.\footnote{A epoch is called radiation dominated if the radiation 
density dominates all other terms in the Friedmann equation. The terms 
``matter dominated epoch'' or ``$\Lambda$--dominated'' epoch are 
explained in the same manner.} One can see, that even in the matter 
dominated epoch the scaling behaviour can not be found. This result can 
easily be understood: As emphasized by Martins 
\& Shellard, the transition from the scaling behaviour in the radiation 
dominated epoch to the scaling in the matter dominated epoch is 
long \cite{martinsshellard}. In the low density models 1, 3 and 4, 
the matter dominated epoch is too short for the string 
network to settle down towards the scaling behaviour, before the curvature 
term or the $\Lambda$--term dominates the expansion of the universe. 

Of course, in models with $\Omega_0 \approx 0.9$ the string network 
reaches scaling in the matter dominated epoch. However, such a model 
is indistinguishable from an Einstein--de Sitter model from the 
viewpoint of structure formation.

Very important for structure formation is the transition--regime 
between the radiation to the matter dominated epoch. We refer to 
the contribution by Richard Battye in these proceedings. 

In the cosmic string theory three mechanism for structure 
formation are possible: accretion on loops, wake formation due to 
fast moving strings and filamentary accretion onto slow moving 
strings, see e.g. \cite{vilenkinshellard, brandenberger, brandenberger2}. 
We us restrict here only on one of these possibilities: the formation of 
wakes, which is suggested by detailed simulations of 
the network evolution, see the discussion in \cite{vilenkinshellard} and 
references therein. 

\section{The peculiar velocity field in cosmic string models}
Every model of structure formation is confronted with the observed 
pro\-per\-ties of the peculiar velocity field of the galaxies. An 
astonishing observed feature is that very large volumes ($R\geq 20$ Mpc) 
flow coherently through space (bulk flows) with large velocities
($v_{\rm bulk}\geq 100$ km/s, depending on the scale), suggesting 
the existence of structures on very large scales, see e.g. 
\cite{dekel} and references therein. 

Here, we calculate the bulk flows predicted from the cosmic string 
theory. In cosmic string theory, the peculiar velocity field of matter 
originates due to the conical space--time structure of a cosmic string. 
If a cosmic string passes a particle it will get a velocity kick in the 
direction swept out by the string \cite{vilenkinshellard}. 
The velocity kick is given by
\begin{equation}\label{kick}
v_{\rm kick} = 4 \pi G \mu \gamma_{s} v_{s} f = 3.8 (G\mu)_{6} (\gamma_{s}
v_{s})f \mbox{ }{\rm km/s},
\end{equation}
with 
\begin{equation}
f =  1 + \frac{1}{2v_{s}^{2}\gamma_{s}^{2}}\left(1 
       - \frac{T}{\mu} \right).
\end{equation}
The term $f$ corresponds to the small scale structure along the 
string, where $\mu$ is the effective mass per unit length along 
the string and $T$ is the string tension. $v_s$ is the string velocity 
and $\gamma_s$ is the corresponding Lorentz--factor.

On this basis it is easy to calculate 
the peculiar velocity field: the peculiar velocity $\vec v$ of a given 
volume is the sum of the influence of many strings, which could influence 
the volume:
\begin{equation}
\vec v = v_{\rm kick} \sum {\cal A}_i \vec{k}_{i},
\end{equation}
where $v_{\rm kick}$ is the value of the kick velocity due to a string, 
${\cal A}_i$ is the amplification factor, which describes the amplification 
of the velocity kick due to gravitational instability and $\vec{k}_i$ is 
an unit vector pointing into the direction of the area 
swept out by string $i$. 
The numbers of strings which could influence the volume are calculated 
with the modified one--scale model by 
Martins \& Shellard \cite{martinsshellard}.
The amplification factors can be calculated with the Zeldovich--approximation 
and we assume that the strings are uncorrelated and therefore 
$< \vec{k}_i \cdot \vec{k}_j > = \delta_{ij}$, where $\delta_{ij}$ is 
the Kronecker symbol \cite{vdbruck2}.
\begin{figure}[htb]
\begin{center}
    \leavevmode
    \epsfxsize=8.9cm
    \epsffile{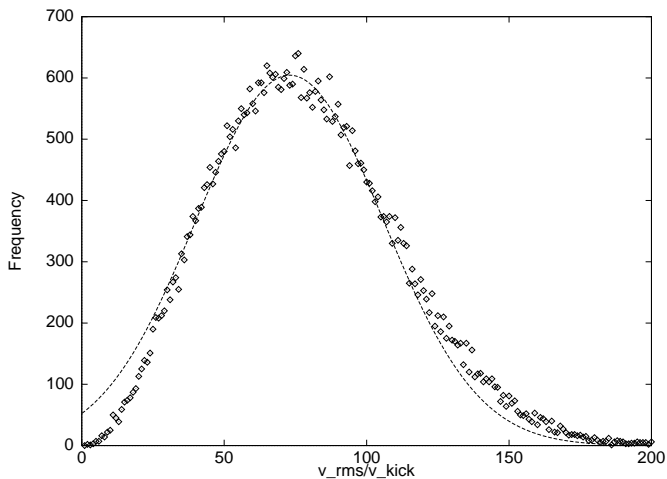}\vspace{0.5cm}
\caption{Probability function of the peculiar--velocity field $v_{\rm rms}$
(in units of the kick velocity $v_{\rm kick}$) 
in the flat $\Lambda$--dominated model (model 4). After 50000 realisations the 
simulations produce an Maxwellian distribution, which shows the 
Gaussian character of each component of the peculiar velocity field, 
adapted from \cite{vdbruck2}. The dotted line is a gaussian curve to calculate 
the expected bulk flow (eq.10).} 
\end{center}
\end{figure}

We note that the peculiar velocity field can be described by a 
gaussian distribution (i.e. each component of $\vec v$ is a random 
variable with a gaussian distribution). This is consistent with the results 
by other authors (e.g. \cite{moessner2}). The probability distribution of the 
rms velocity $v_{\rm rms}$ is plotted in Fig.3 for the 
flat $\Lambda$--dominated model (model 4) and in Fig.4 for the open model 
(model 3). For the other models we observe the same behaviour. 

\begin{figure}[htb]
\begin{center}
    \leavevmode
    \epsfxsize=8.9cm
    \epsffile{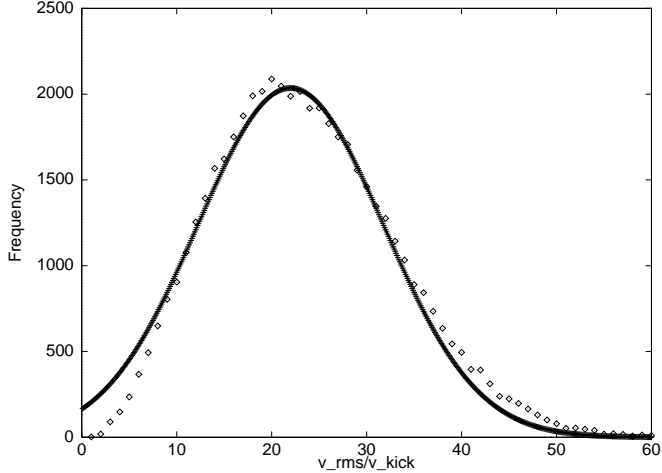}\vspace{0.5cm}
\caption{The same as in Fig.3 but here for the open model. The solid line 
is a gaussian curve to calculate the expected bulk flow (eq.11).}
\end{center}
\end{figure}
 
We calculate $v_{\rm rms}$ from the probability distribution 
by fitting a gaussian (see Fig.3 and Fig.4). In the models (Table 1)
we obtain a peculiar velocity at a 
scale corresponding to $L(t_{eq})$ given by (the length scales 
are calculated with $H_0 = 100$km/(s$\cdot$Mpc)) \cite{vdbruck2}:
\begin{eqnarray}\label{result1}
v_{pec}(L_{eq}\approx 70\mbox{ }{\rm Mpc})_{\rm closed} \\ \nonumber
 = (460 \pm 200)(G\mu)_{6} (\gamma_{s}v_{s})f\mbox{ } {\rm km/s},
\end{eqnarray}
\begin{eqnarray}
v_{pec}(L_{eq}\approx 1\mbox{ }{\rm Mpc})_{\rm EdS} \\ \nonumber
= (1740 \pm 760)(G\mu)_{6} (\gamma_{s}v_{s})f\mbox{ }{\rm km/s},
\end{eqnarray}
\begin{eqnarray}
v_{pec}(L_{eq}\approx 10\mbox{ }{\rm Mpc})_{\lambda,{\rm flat}} \\ \nonumber
= (280 \pm 120)(G\mu)_{6} (\gamma_{s}v_{s})f\mbox{ }{\rm km/s},
\end{eqnarray}
\begin{eqnarray}\label{result2}
v_{pec}(L_{eq}\approx 10\mbox{ }{\rm Mpc})_{\rm open} \\ \nonumber
= (80 \pm 35)(G\mu)_{6} (\gamma_{s}v_{s})f\mbox{ } {\rm km/s}.
\end{eqnarray}
The length scale corresponding to the time $t_{eq}$ is set to be  
$0.1 H^{-1}(t_{eq})$ \cite{vachaspati}:
\begin{equation}
L_{eq} \approx 1.1 \frac{1}{\Omega_{0}} h_{0}^{-2} {\rm Mpc}
\end{equation}

\begin{figure}[htb]
 \label{bulk}
 \begin{center}
  \leavevmode
   \epsfxsize=8.5cm
   \epsfysize=6cm
  \epsffile{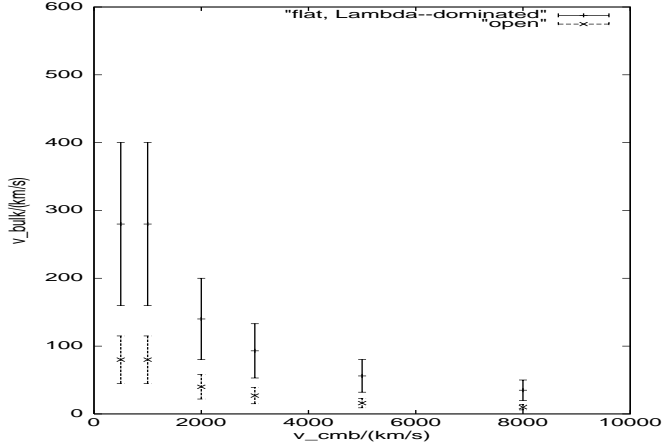}\vspace{0.5cm}
\caption{Expected bulk flows $v_{\rm bulk}$ in cosmic string theory 
as a function of redshift in a flat, 
$\Lambda$--dominated model (model 4) and an open model (model 3). 
Here we assume $(G\mu)_{6} (\gamma_{s}v_{s})f \approx 1$. The curves 
would shift if this quantity differs from unity (see text for details). 
One can see that cosmic strings in open or flat cosmological models 
predict a ``quiet'' peculiar velocity field at large scales. Note that 
in the closed model 1 this is not the case: the peculiar velocities are 
constant over a scale of $\leq 10000$ km/s and of the order 
$450 \pm 200$ km/s.}
\end{center}
\end{figure}

Because the primordial spectrum of the density 
fluctuations in cosmic string theories is nearly of a 
Harrison--Zeldovich type, the peculiar velocity 
field is roughly proportional to $L^{-1}$ (see Fig.5).
From this follows, for example in the open model (model 3) 
with $\Omega_0=0.1$ at a 
scale of $L=50 h^{-1}$ Mpc, a bulk velocity around 20 km/s, whereas 
in the flat $\Lambda$--dominated model (model 4) the velocity is 
around 70 km/s. 
This is inconsistent with the POTENT--results (see e.g. \cite{dekel}), 
as far as $(G\mu)_{6} (\gamma_{s}v_{s})f$ is of the order 1, 
and with the Lauer--Postman result \cite{lauer}, which suggest bulk flows 
on large scales. These results are, however, consistent with the 
observed values obtained by Giovanelli et al. \cite{giov}, which suggest 
a much more quiescent peculiar velocity field. The closed model fits the 
POTENT results well at each length scales. 

The fact, that many cosmic strings influence the matter in the universe 
and the peculiar velocity field led us to ask the question if we can 
then estimate the density parameter from the comparison between the 
density field and the peculiar velocities. Such investigations are 
based on the equation 
\begin{equation} \label{potent}
\nabla \cdot {\bf v} = -\beta H \delta,
\end{equation}
with $\beta = \Omega_0^{0.6}$, 
which holds in linear perturbation theory for an irrotational 
fluid\footnote{Cosmic strings might generate vorticity also at large 
scales. However, these quickly decay by cosmic expansion ($\propto R^{-2}$) 
so that we can neglect it at the current epoch.}. However, even if there 
is a relation between velocity and density field, eq.(\ref{potent}) is 
derived from the continuity equation and one can give examples in which 
an overestimated $\Omega_0$ could be derived from eq.(\ref{potent}) 
(see the discussion for the case of the explosion scenario in \cite{babul}). 
In the case of strings one can easily show that there is an connection 
between density and peculiar velocity field of the form
\begin{equation} \label{strings}
\nabla \cdot {\bf v} = -\beta_{\rm eff} H \delta,
\end{equation}
with
\begin{equation}
\beta_{\rm eff}=\frac{\sum_{i,j}\delta_{i,j}
\beta_{{\rm eff},i,j}}{\sum_{i,j}\delta_{i,j}}.
\end{equation}
Here, $\delta_{i,j}$ is the density fluctuation created by the string $i,j$ 
and $\beta_{{\rm eff},i,j}$ is an ``effective'' value which connects the 
density and the velocity field of the string $i,j$, see \cite{vdbruck2}. 
In general, $\beta_{\rm eff}$ should be different from $\Omega_0^{0.6}$. 
This let us to conclude that in general on can not obtain $\Omega_0$ from 
a comparison of the density and the velocity field if cosmic strings 
where responsible for structure formation \cite{vdbruck2}. However, 
important for this conclusion is the knowledge of the length scale at 
which non--gaussianity becomes important\footnote{I am grateful to 
Pedro Avelino for pointing out this to me.}. One has to find this 
length scale with numerical simulations. 

\section{Non--gaussianity in cosmic string models}
Avelino et al. investigated this question for CDM and HDM models in 
detail \cite{avelino2}. They calculated 
the non--gaussian scale $R_{\rm ng}$, below which the non--gaussian 
statistics of the density field is 
very important. For scales larger than $R_{\rm ng}$, the statistics 
of the density field can be described by a gaussian distribution to 
high accuracy. We discuss here the case of CDM.

In order to find $R_{\rm ng}$, Avelino et al. calcuated the 
skewness and kurtosis of the smoothed density field (i.e. of the 
probability density function) \cite{avelino2}. Because the 
kurtosis and the skewness 
are zero for a gaussian distribution, these statistics give a hint for 
non--gaussianity if they are non--zero. Avelino et al. also calculated 
the genus curves of the density field, which is also a good indicator 
for the topology of the density field.

\begin{figure}[htb]
 \begin{center}
  \leavevmode
   \epsfxsize=8.5cm
   \epsfysize=6cm
  \epsffile{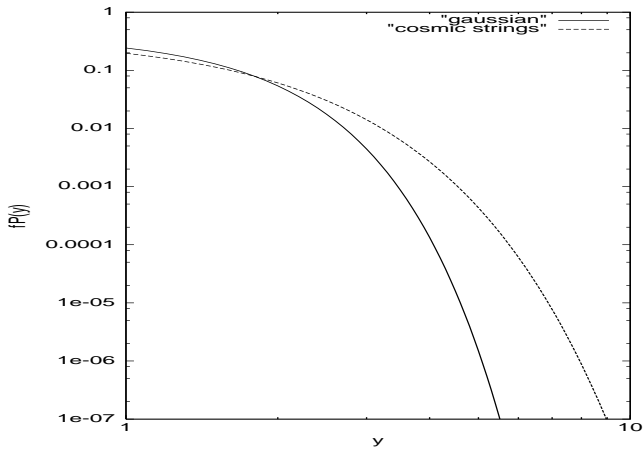}\vspace{0.2cm}
\caption{The tails of the effective probability function $fP(y)$ 
for gaussian and cosmic string models on the scale $R_{\rm ng}$. 
The cosmic string data are taken from \cite{avelino2}. 
$y=\delta/\sigma_R$, where $\delta$ is 
the density contrast and $\sigma_R$ is the mass fluctuation at the 
scale $R_{\rm ng}$, see the text for details.}
\end{center}
\end{figure}

As a result they found 
$R_{\rm ng} = 1.5 (\Omega_0 h^2)^{-1}$ Mpc. For models with 
$\Omega_0 \approx 1$ this length scale is very small compared to 
50 Mpc at which large scale flows are observed. Therefore, for these 
models the density and peculiar velocity field are related by equation 
(\ref{potent}). In fact, only for models with $\Omega_0 h^2 \leq 0.1$ 
the scale is large enough that a significant departure from 
eq.(\ref{potent}) is important (for example model 1). Then 
eq.(\ref{strings}) holds. Cosmic strings in
such very low density models might be ruled out by the 
correlation analysis in Fourier space, see Stirling \& Peacock \cite{peacock}. 
They found from the combined QDOT and 1.2--Jy IRAS galaxy survey 
that the modes with wavelength $\lambda > 30 h^{-1}$ Mpc are in 
good agreement with a gaussian distribution. While this is a provisional 
conclusion, with future observations this test will place limits on 
the degree of non--gaussianity in the density fluctuation field on 
small and large scales. For another discussion on non--gaussianity 
see \cite{mitsouras}.

In low--density models with $\Omega_0 h = 0.1 ... 0.2$ the length scale 
$R_{\rm ng}$ is about 10 Mpc. From this length scale clusters collapse. 
Therefore, different predictions of the number density of cluster of 
galaxies are expected. In fact, clusters are used to test cosmological 
models, see e.g. \cite{bartlett} and references therein. 
Here we use the Press--Schechter method to caluclate the abundances 
of clusters in cosmic string theory. Because of non--gaussianity we
have to use a modified approach, see e.g. \cite{chiu}. 

We consider the smoothed density field $\delta$. The probability 
density function $p(\delta)$ can be described by a rescaled function 
$P(y)$ with $y=\delta/\sigma_R$, where $\sigma_R$ is the mass fluctuation at 
the scale $R$. We assume $p(\delta) = P(y)/\sigma_R$ 
which should hold in the theories considered here and on the relevant 
length scales. The probability density function $P(y)$ is shown 
for cosmic string models and gaussian models in Fig.6.

In the Press--Schechter formalism the number density $n(R)$ of 
collapsed objects (collapsed from a radius $R$) is given by
\begin{equation}
n(R)dR = \frac{3}{4 \pi R^3} \left| \int_{y_c}^{\infty} fP(y) dy \right|dR,
\end{equation}
where $f=1/\int_0^{\infty} P(y) dy$ and $y_c = \delta_c/\sigma_R$. Here 
$\delta_c$ is the critical density contrast for collapse. We assume the 
spherical collapse model with $\delta_c=1.68$, because this quantity depends 
only very weakly on the cosmological model \cite{eke}. From the power 
spectrum we calculated $\sigma_R$. It can be shown that 
$\sigma_R \propto R^{-\alpha}$, where $\alpha \approx 1$. Our results for 
the range 1 keV to 10 keV depend not very strongly on 
the slope of the power spectrum if $\alpha = 0.75...1.25$, 
which was also observed by others \cite{eke}. We give details of our 
calculations elsewhere \cite{vdbfuture}.

\begin{figure}[htb]
 \begin{center}
  \leavevmode
   \epsfxsize=9cm
   \epsfysize=7cm
  \epsffile{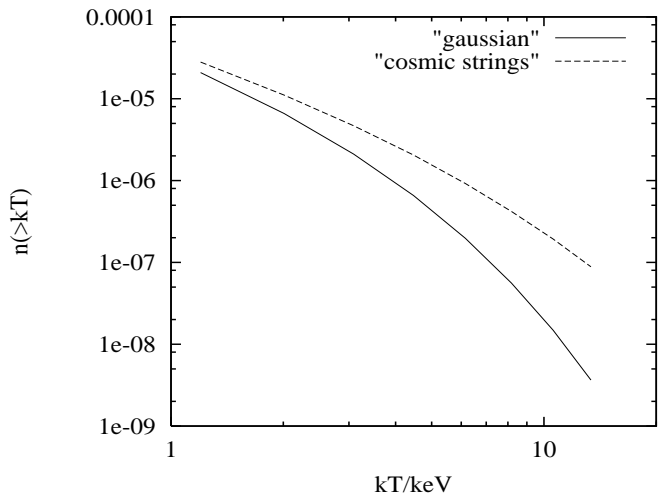}\vspace{0.5cm}
\caption{Number densities of clusters of galaxies as a function of the 
temperature $kT$ in a gaussian and 
a cosmic string model. The cosmological parameters are $\Omega_0 = 0.3$, 
$\lambda_0=0.7$, $h=0.7$ and $\sigma_8=0.75$.}
\end{center}
\end{figure}

Because the length scale $R$ from which a cluster collapsed is not an 
observable, we have to relate $R$ with the cluster mass $M$ or the X--ray 
temperature of the cluster gas. Here we calculate the temperature 
function and assume a spherical cluster collapse. The resulting 
cluster is described by an isothermal sphere with a radius of 
virialisation equal to half its maximum expansion radius (e.g. \cite{white}). 
Then, for a hydrogen mass fraction of 0.76, 
\begin{equation}
kT = 8.6 \mbox{ }{\rm keV} \mbox{ }\Omega_0^{2/3} 
     \left(\frac{R}{10 h^{-1} {\rm Mpc}}\right)^{2}
     \left(\frac{\Omega_0}{\Omega(z_f)}\right)^{1/3} 
     \left[\frac{\Delta_c}{178} \right]^{1/3},
\end{equation}
where $z_f$ is the redshift of collapse, which we calculate with the 
spherical collapse model. 
We calculate the number density $n(>kT)$ of clusters which X--ray 
temperatures greater than $kT$ (temperature function). 

\begin{figure}[htb]
 \begin{center}
  \leavevmode
   \epsfxsize=9cm
   \epsfysize=7cm
  \epsffile{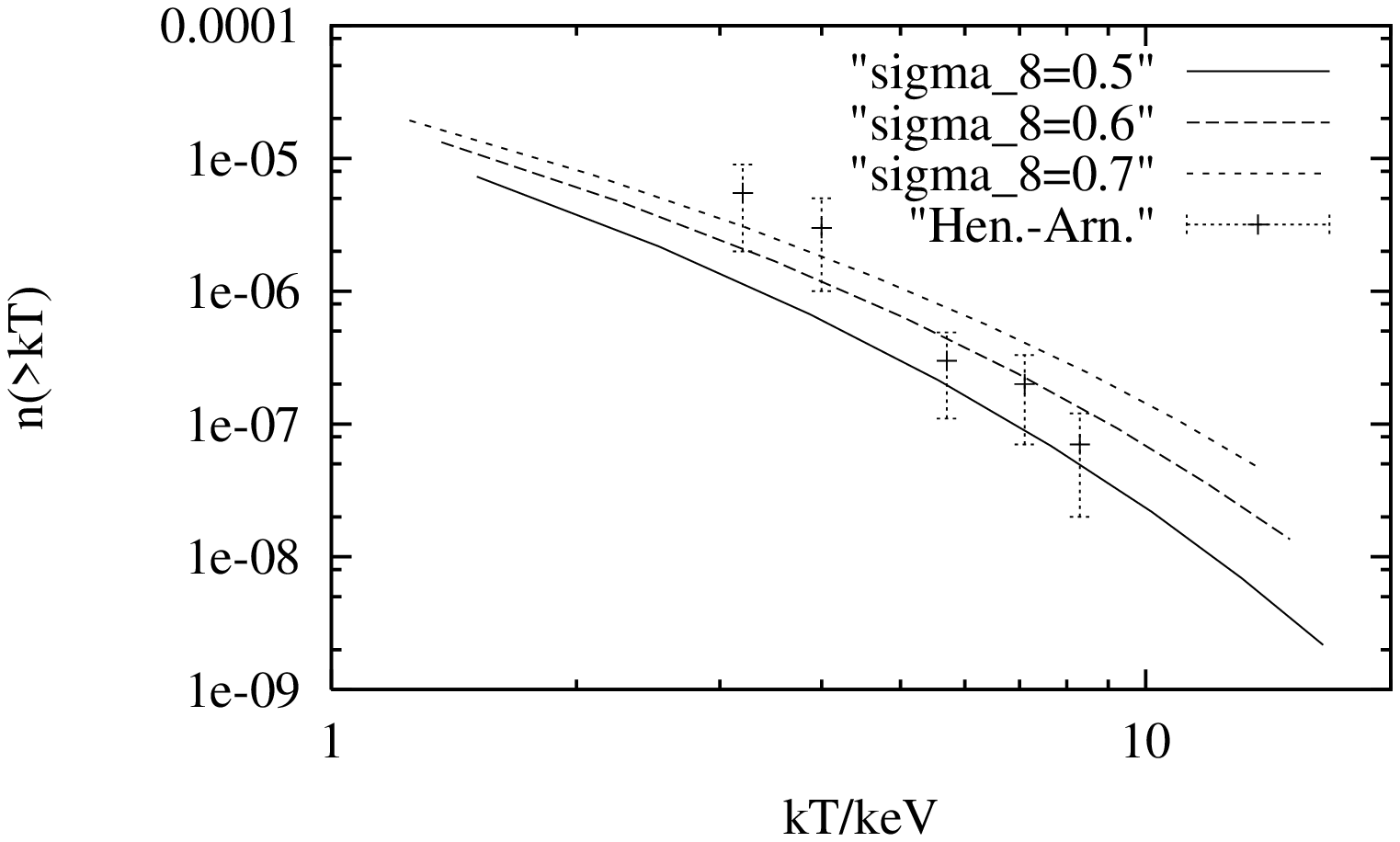}\vspace{0.5cm}
\caption{Number density of clusters of galaxies in cosmic string models with 
$\Omega_0 = 0.3$, $\lambda_0=0.7$, $h=0.7$ and different $\sigma_8$ 
as a function of the X--ray temperature. Shown are also the temperatur 
function according to Henry \& Arnaud \cite{henry1}.}
\end{center}
\end{figure}

Because high density peaks are much more common in cosmic string 
models as in gaussian models, for a given density parameter, cosmological 
constant, Hubble parameter and mass fluctuation at 8 $h^{-1}$ Mpc, 
cosmic strings predict more clusters with high X--ray temperatures 
as compared to gaussian models. 
This can be seen in Fig.7. If we change $\sigma_8$ the temperature function 
changes (Fig.8). Also shown are the data by 
Henry \& Arnaud (1991) (the vertical bars with the range of uncertainty). 
We found that $\sigma_8$ should lie in the range $0.6-0.7$ to fit the 
data (for a flat model with $\Omega_0 = 0.3$). The reader is referred to our 
forthcoming work, where more data will be used and where the errors, 
which could enter the calculations, are discussed in detail \cite{vdbfuture}.

Similar to gaussian models we find that the temperature function (and also 
the mass function) are good discriminators for non--gaussian models. 
We refer to our future work, where these functions and cluster 
evolution will be discussed in some depth \cite{vdbfuture}.

We are not able to discuss cluster abundances in the baryonic model 1 of 
Tab.1 because we don't know the probability function on the scale 
relevant for cluster formation. 

In conclusion we point out, that several uncertainties enter the analyis:
\begin{itemize}
\item We concentrate here only on the wakes. How important are cosmic string 
loops for structure formation? Is filament formation due to slow moving 
strings also important? If these mechanism turn out to be important, 
the probability density function of the density field will certainly 
change. 
\item How does the cluster gas evolve and what is its 
impact on the temperature function (e.g. \cite{pen})? 
Here, an detailed analysis of the formation and evolution of wakes 
seems necessary, see \cite{sornborger,wandelt}. 
\item How good is the spherical collapse model? What is the typical 
formation redshift of a cluster?
\item Is the Press--Schechter formalism good enough do describe 
the statistics of high--density peaks in non--gaussian models? 
\end{itemize}
At the present time we are far from conclusive answers for the questions 
above. 

The case for HDM models + cosmic strings is not discussed here. But the 
simulations suggest that on cluster scales the probability density 
function is gaussian in these models and cosmic strings may have 
the same problems as in ``usual'' gaussian models 
\cite{balland}\footnote{However, see \cite{moessner}, for a discussion 
of the formation of high redshift objects in cosmic strings + HDM.}. 
Note, however, that it is important to know which kind of 
perturbations are important in cosmic string theory: wakes, filaments 
and/or loops are possible, for a discussion 
see e.g. \cite{brandenberger2,zanchin}.

\section{Discussion}
We have discussed some aspects of the cosmic string theory of structure 
formation. Our results suggest that low density models are very 
encouraging for cosmic string theory. Problems might arise if the 
peculiar velocity field is observed to be of the order of 300 km/s 
at large scales ($cz \geq 3000$ km/s). Then only models with 
$\Omega_0 h \leq 0.1$ and a cosmological constant 
could produce such large flows at these scales 
(if $(G\mu)_{6} (\gamma_{s}v_{s})f \approx 1$). Such models predict 
non--gaussian features at these scales and could be constrained 
with topological statistics, see e.g. \cite{mitsouras} or with 
correlation analysis in Fourier space, see e.g. \cite{peacock}.

Further observational efforts 
will help to decide if cosmic strings can be reponsible for structure 
formation. In particular if the peculiar velocity field is better constrained 
from observations at large scales and if we understand cluster 
formation and evolution more thoroughly, we should able to test these 
theories apart from CMB observations and matter distribution. 

\vspace{1cm}

{\bf Acknowledgements:} I would like to thank 
Prof. H.V. Klapdor--Kleingrothaus and Laura Baudis for organizing 
an interesting conference. I thank Richard Battye for useful discussions 
on the peculiar velocity field and August Evrard 
for useful discussions during the conference. Gratefully acknowledged are 
discussions with Paul Shellard and in particular with Pedro Avelino during 
a visit at DAMTP/Cambridge. I am also grateful to Proty Wu for submitting 
the data of the probability function and to Wolfgang Priester and B\"ulent 
Uyan{\i}ker for a careful reading of the manuscript. This work was supported 
by the Deutsche Forschungsgemeinschaft (DFG) and the Max Planck--Gesellschaft.


\begin{thebibliography}{15}
\bibitem{avelino1} Avelino, P.P., Caldwell, R.R., Martins, C.J.A.P., 1997, 
Phys.Rev.D {\bf 56}, 4568 
\bibitem{avelino2} Avelino, P.P., Shellard, E.P.S., Wu, J.H.P., Allen, B., 
astro--ph/9803120, to appear in Astrophys. Journ. Lett. 
\bibitem{babul} Babul, A., Weinberg, D. H., Dekel, A., Ostriker, J.P., 
1994, ApJ {\bf 427}, 1
\bibitem{balland} Balland, C., Blanchard, A., 1995, A\&A {\bf 298}, 323
\bibitem{bartlett} Bartlett, J.G., 1997, in: {\it From quantum fluctuations 
to cosmological structures}, eds. Valls--Gabaud, D., Hendry, M., Molaro, P. 
and Chamcha, K., ASP Conference Series {\bf 126}, p.365
\bibitem{botun} Bothun, G., 1998, {\it Modern cosmological observations and 
problems}, Taylor \& Francis, London
\bibitem{brandenberger} Brandenberger, R., 1994, Int. Journ. Mod. 
Phys. A {\bf 9}, 2117
\bibitem{brandenberger2} Brandenberger, R., 1991, Phys. Scripta T{\bf36}, 
114
\bibitem{chiu} Chiu, W.A., Ostriker, J.P., Strauss, M.A., 1998, 
ApJ {\bf 494}, 479
\bibitem{dekel} Dekel, A., 1997, in: {\it Galaxy scaling relations: origins, 
evolution and applications}, ed. L. da Costa, Springer
\bibitem{eke} Eke, V.R., Cole, S., Frenk, C.S., 1996, MNRAS {\bf 282}, 263
\bibitem{hindmarsh} Hindmarsh, M, Kibble, T.W.B., 1995, 
Rep.Prog.Phys. {\bf 58}, 477 
\bibitem{kolbturner} Kolb, E., Turner, M.S., 1991, {\it The early universe}, 
Addison Wesley
\bibitem{lauer} Lauer, T., Postman, M., 1994, ApJ {\bf 425}, 418
\bibitem{martinsshellard} Martins, C.J.A.P., Shellard, E.P.S., 1996, 
Phys.Rev.D {\bf 54}, 2535
\bibitem{giov} Giovanelli R., et al., 1996, ApJ {\bf 464}, L99
\bibitem{henry1} Henry, J.P., Arnaud, K.A., 1991, ApJ {\bf 372}, 410 
\bibitem{mitsouras} Mitsouras, D., Brandenberger, R., Hickson, P., 1998,
astro--ph/9806360, submitted to MNRAS
\bibitem{moessner2} Moessner, R., 1995, MNRAS {\bf 277}, 927
\bibitem{moessner} Moessner, R., Brandenberger, R., 1996, MNRAS {\bf 280}, 797
\bibitem{peacock} Stirling, A.J., Peacock, J.A., 1996, MNRAS {\bf 283}, 99
\bibitem{pen} Pen, U.--L., 1998, ApJ {\bf 498}, 60
\bibitem{sornborger} Sornborger, A., Brandenberger, R., Fryxell, B., 
Olson, K., 1997, ApJ {\bf 482}, 22
\bibitem{vachaspati} Vachaspati, T., 1992, Phys. Lett. B {\bf 282}, 305
\bibitem{vdbruck1} van de Bruck, C., 1998, Phys.Rev.D {\bf 57}, 1306
\bibitem{vdbruck2} van de Bruck, C., 1998, Phys.Rev.D {\bf 57}, 4663
\bibitem{vdbfuture} van de Bruck, C., 1998, in preparation
\bibitem{vilenkinshellard} Vilenkin, A., Shellard, E.P.S., 1994, {\it Cosmic 
strings and other defects}, Cambridge University Press
\bibitem{wandelt} Wandelt, B.D., 1998, ApJ {\bf 503}, 67
\bibitem{white} White, S.D.M., Efstathiou, G., Frenk, C.S., 1993, 
MNRAS {\bf 262}, 1023
\bibitem{zanchin} Zanchin, V., Lima, J.A.S., Brandenberger, R., 1996, 
Phys.Rev.D {\bf 54}, 7129
\end{thebibliography}
\end{document}